RESEARCH ARTICLE

# Mechanisms underlying different onset patterns of focal seizures

**Yujiang Wang**[1,2,3]*, **Andrew J Trevelyan**[2], **Antonio Valentin**[4], **Gonzalo Alarcon**[4,5], **Peter N Taylor**[1,2,3], **Marcus Kaiser**[1,2]

**1** Interdisciplinary Computing and Complex BioSystems (ICOS) research group, School of Computing Science, Newcastle University, Newcastle upon Tyne, United Kingdom, **2** Institute of Neuroscience, Newcastle University, Newcastle upon Tyne, United Kingdom, **3** Institute of Neurology, University College London, London, United Kingdom, **4** Department of Basic and Clinical Neuroscience, Institute of Psychiatry, Psychology and Neuroscience, King's College London, London, United Kingdom, **5** Comprehensive Epilepsy Center, Neuroscience Institute, Academic Health Systems, Hamad Medical Corporation, Doha, Qatar

* Yujiang.Wang@Newcastle.ac.uk

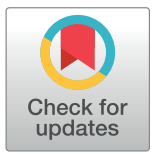

## Abstract

Focal seizures are episodes of pathological brain activity that appear to arise from a localised area of the brain. The onset patterns of focal seizure activity have been studied intensively, and they have largely been distinguished into two types—low amplitude fast oscillations (LAF), or high amplitude spikes (HAS). Here we explore whether these two patterns arise from fundamentally different mechanisms. Here, we use a previously established computational model of neocortical tissue, and validate it as an adequate model using clinical recordings of focal seizures. We then reproduce the two onset patterns in their most defining properties and investigate the possible mechanisms underlying the different focal seizure onset patterns in the model. We show that the two patterns are associated with different mechanisms at the spatial scale of a single ECoG electrode. The LAF onset is initiated by independent patches of localised activity, which slowly invade the surrounding tissue and coalesce over time. In contrast, the HAS onset is a global, systemic transition to a coexisting seizure state triggered by a local event. We find that such a global transition is enabled by an increase in the excitability of the "healthy" surrounding tissue, which by itself does not generate seizures, but can support seizure activity when incited. In our simulations, the difference in surrounding tissue excitability also offers a simple explanation of the clinically reported difference in surgical outcomes. Finally, we demonstrate in the model how changes in tissue excitability could be elucidated, in principle, using active stimulation. Taken together, our modelling results suggest that the excitability of the tissue surrounding the seizure core may play a determining role in the seizure onset pattern, as well as in the surgical outcome.

## Author summary

Much attention has been devoted to the mechanisms underlying epileptic seizures. However, so far, the morphology of how seizures start on electrographic recordings project (http://www.greenbrainproject.org) funded through Engineering and Physical Sciences Research Council (EPSRC) (EP/K026992/1), and the Portabolomics Project funded through Engineering and Physical Sciences Research Council (EPSRC) (EP/N031962/1). The funders had no role in study design, data collection and analysis, decision to publish, or preparation of the manuscript.

**Competing interests:** The authors have declared that no competing interests exist.

(i.e. the seizure onset pattern) has been neglected as a potential indicator of the underlying dynamic mechanism. In this work, we take a spatio-temporal modelling approach to reproduce and understand two major seizure onset patterns. We find that it is not necessarily the initiation in the seizure core that determines onset pattern, but that the excitability of the surrounding "healthy" tissue plays a pivotal role in how the seizure onset appears. In agreement with previous computational modelling work, we hypothesise that indeed the two patterns could arise due to different dynamic onset mechanisms, where the surround excitability differs fundamentally in their dynamic properties. Our hypothesis indeed also offers a simple explanation for the clinically reported difference in surgical outcome for the two onset patterns. As an outlook, we also propose a possible way to track such changes in the surround excitability in a proof-of-principle computational demonstration.

## Introduction

Focal seizures are episodes of highly disruptive brain activity, which are considered to arise from local sites of pathological abnormality in the brain. Identification of the precise focal origin in a given patient is crucial for the clinical management of their epilepsy. Various clinical evidence can point to the nature and site of the origin, but critical amongst these are the data recorded by electroencephalography (EEG) or by the invasive alternative of electrocorticography (ECoG). Previous clinical studies have suggested that focal seizures could have different onset mechanisms [1–4]. Related to this, there is a clear heterogeneity in clinical outcomes: only a subset of patients responds to medications [5], or surgical removal of the presumed epileptic focus [6–8].

In terms of the appearance and morphology of focal seizure activity, generally two types of common onset patterns are reported: The low amplitude fast (LAF) activity, which is characterized by oscillations in the beta to gamma range of initially low amplitude that slowly increases as the seizure progresses; and the high amplitude slow (HAS) activity, which is generally described as a slower oscillation below the alpha range with a high amplitude at onset [9–18]. Depending on the study, and subtype of epilepsy, finer distinctions in the onset pattern have been characterized, and different quantitative measures have been used to categorize the onset patterns. For example, many studies use a cut-off frequency in the alpha band to distinguish between LAF other onset patterns [10, 11, 16, 17], but the choice generally varied between 8 Hz and 20 Hz.

The majority of studies find that the LAF pattern is most frequently occurring pattern [10, 14, 16–18]. Intriguingly, despite different categorization criteria, and subtypes of epilepsy included, it appears that the LAF pattern is associated with a good surgical outcome when removed [9, 10, 12, 14, 16, 18]. Depending on the study, and subtype of epilepsy considered, a potentially worse surgical outcome is implicated for the HAS compared to the LAF pattern [10, 12, 14, 16]. However, the literature is not entirely consistent; such a state usually reflects difficulties arising from limited clinical information, perhaps requiring more precision either in the classification of the diagnosis, or of the nature of the onset pattern [19, 20].

Furthermore, the LAF pattern is associated with a larger seizure onset zone (in both temporal and extratemporal lobe seizures), which is measured by the number of contacts first showing epileptic activity [11, 17]. Additionally, the cortical excitability as measured by cortico-cortical evoked potentials (CCEPs) is lower in the seizure onset zone of the LAF pattern compared to the HAS pattern [15]. The evidence presented here indicates distinct mechanisms generating the different onset patterns. However, to our knowledge, no spatio-temporal interpretation exists so far explaining the basis of these distinct mechanisms.

To provide a mechanistic explanation, we turn to our previously suggested theoretical framework, which demonstrated the possible heterogeneity of onset mechanisms in focal seizures [21]. Our model utilised established concepts of increased excitability [2, 22], impaired inhibitory restraints [23–25], bistability and triggers [26–28] to provide a classification of dynamic mechanisms of focal seizure onset. We now show that two of the distinct classes of dynamical mechanisms described in our earlier study appear to correspond to the clinical LAF and HAS patterns.

## Results

### Modelling low and high amplitude onset patterns

To begin the investigation of seizure onset patterns in the model, we first show that the model is capable of reproducing the clinical observations of low amplitude fast (LAF), and high amplitude slow (HAS) patterns. As the clinical categorisation of the onset patterns differ considerably across different studies, we focused on the amplitude of the pattern in our model as a main aspect to reproduce. Most studies agree that the LAF pattern is a very low amplitude activity compared to other patterns, which gradually grows over time. Interestingly, as we shall see, the LAF pattern is intrinsically associated with a higher frequency of oscillation than the HAS pattern in our model.

The LAF onset is shown in Fig 1A. In agreement with the clinical recording, the seizure onset on the simulated ECoG electrode (which is placed over the entire simulated cortical sheet) in the model arises out of the background state with low amplitude fast oscillations (~14 HZ). Spatio-temporally, the LAF activity on the simulated ECoG is caused by small patches of microdomains displaying localised oscillatory microseizure activity. As the simulated seizure progresses, more microdomains become involved, and the surrounding tissue is recruited slowly. In other words, the microdomains coalesce slowly and form larger contiguous and coherent patches of seizure activity. The amplitude of the simulated ECoG average activity also grows. Eventually, the entire simulated tissue is recruited into the seizure activity. The slowing down in frequency of the simulated ECoG oscillations as the seizure progresses is a direct consequence of the recruitment process. The oscillatory frequency of seizure activity in smaller patches is intrinsically faster than the frequency of a large contiguous piece of tissue.

The high amplitude onset is shown in Fig 1B. Also in agreement with the clinical recording, the seizure starts with a HAS oscillation on the simulated ECoG. The oscillation starts relatively slow (~9 Hz) compared to the LAF pattern. On the spatio-temporal view, it becomes clear that the initiation and subsequent evolution of this HAS pattern is very different from that of the LAF onset in the model. Upon activation of one patch of localised seizure activity, this activity invades the surrounding tissue rapidly as a propagating wavefront. Some long-range activation also appears subsequently through the long-range connections in the model cortical sheet. The entire simulated cortical sheet is fully recruited within 0.8 s. In contrast to the LAF onset patterns, the low frequency at onset of HAS onset patterns remains fairly constant, even throughout the initial evolution phase (first 1-2 seconds), similar to the clinical observation.

We also want to highlight here that the model reproduces not only the initial onset pattern, but to some extent also the evolution of the seizure. For example, the growing amplitude in the LAF pattern is shown in our model, which after several seconds develops into a high-amplitude spiking activity. However, our model does not simulate the termination of the seizure. Hence, we are only proposing to investigate the mechanism of onset and initial evolution of the seizure. The mechanism of the complete seizure development and eventual termination

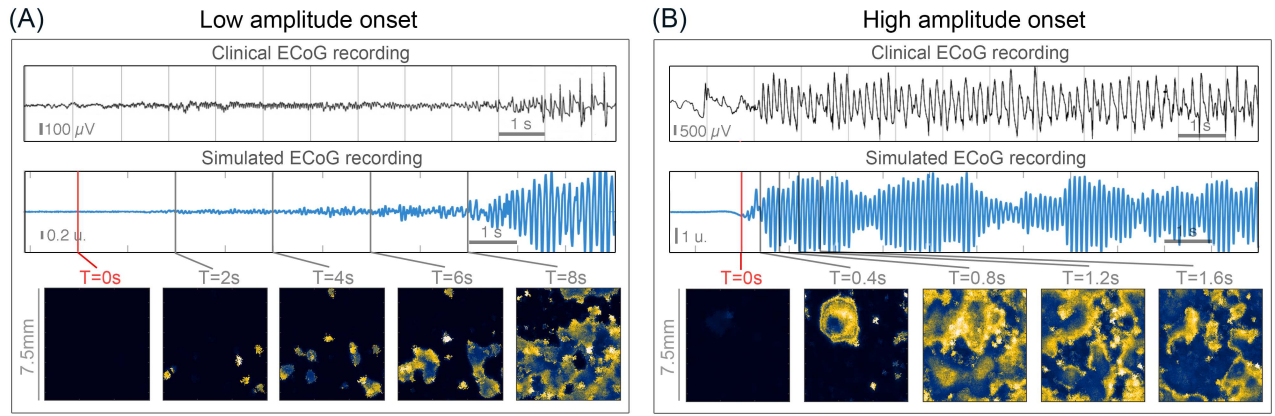

**Fig 1. Modelling low and high amplitude onset patterns. (A)** Low amplitude fast (LAF) onset pattern. Top: example clinical recording of a LAF pattern (onset indicated by red line), visualised from [17]. Middle: Simulated ECoG electrode recording of a LAF pattern. Seizure onset was initiated by activating patches of microseizure activity (marked by red line). Bottom: corresponding snapshots of the simulated activity on the cortical sheet. **(B)** High amplitude slow (HAS) onset pattern. Top: example clinical recording of a HAS pattern (onset indicated by red line), from the iEEG database www.ieeg.org, Study 020, seizure 3. Middle: Simulated ECoG electrode recording of a HAS pattern. Seizure onset was initiated by activating patches of microseizure activity (marked by red line). Bottom: corresponding snapshots of the simulated activity on the cortical sheet. The time scale of the clinical recordings is matched with those of the simulations. The red vertical bar in the simulation indicates T = 0 s. The parameters used can be found in Table 1.

https://doi.org/10.1371/journal.pcbi.1005475.g001

may be driven by processes on longer time scales [29, 30] which we leave as an investigation for a future study.

### High amplitude patterns are associated with an increased excitability of the surround

After confirming that the model is indeed capable of reproducing both onset patterns qualitatively, we turn to a systematic study of the conditions that lead to either pattern.

In our previous work [21], we demonstrated that focal seizure onset can be understood in terms of the spatial regions of seizure activity, and surrounding non-seizing tissue. In particular, we showed the different classes of seizure onset were associated with a different excitability setting of the surrounding territory. Hence, we choose to investigate surround excitability in terms of onset patterns. Here, the term excitability refers to the proximity to the monostable seizure state in the parameter space (see Methods for details). Changes in the excitability levels of the surround can be caused by a range of parameters (e.g. reductions in the inhibitory restraint). In the following, we choose to investigate the parameter $P$ (constant input to the excitatory population). Equivalent results hold for other parameters that induce bistability (e.g. parameter $Q$, see S1 Fig. compared to Fig 2).

We also demonstrated the importance of the spatial organisation of the seizure core in our previous work [21]. In other words, the number and size of microdomains initiating the microseizure activity also influences the overall dynamics at seizure initiation. For example, 10% of minicolumns displaying microseizure activity can either all be spatially arranged into one large patch, or ten separate smaller patches. This example is illustrated in Fig 2C in the top two panels. Therefore, we shall also study the effect of size and density of the microdomains on seizure onset patterns.

Fig 2A shows the resulting scan of seizure amplitude (illustrated in the size of the dots) with respect to the three parameters (i) surround excitability ($P_{surround}$), (ii) total percentage of minicolumns displaying microseizure activity, and (iii) number of microdomains. The high amplitude onset patterns (big dot sizes) are all found for high values of excitability of the

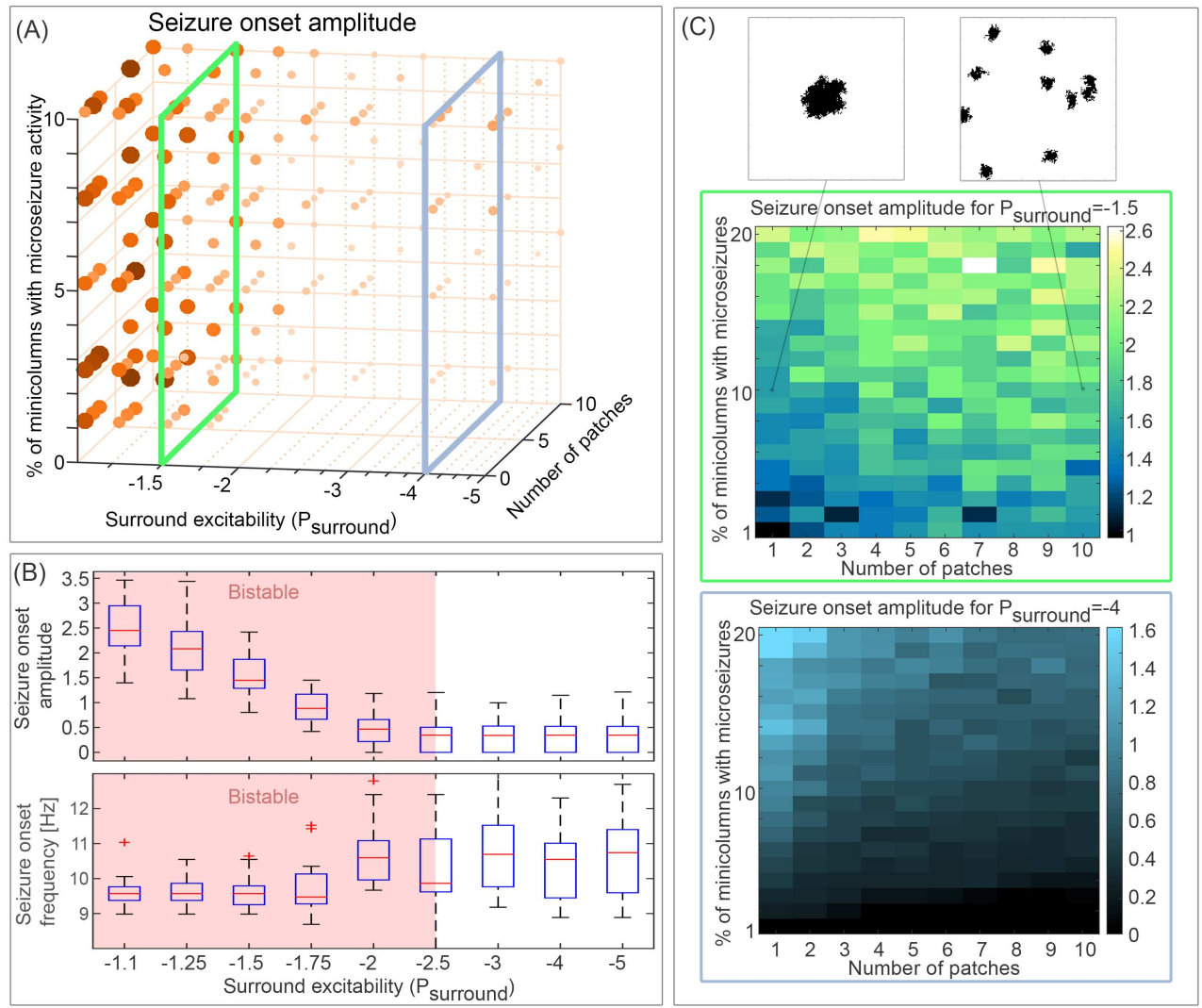

**Fig 2. Effect of different parameters on seizure onset amplitude. (A)** Seizure onset amplitude is shown (as the size and saturation of the circle markers) with respect to the parameters surround excitability ($P_{surround}$), percentage of minicolumns showing microseizure activity, and number of patches. **(B)** Amplitude of onset is shown only for different values of the surround excitability $P_{surround}$. Essentially, the 3D parameter scan in (A) is shown here as a projection onto the parameter $P_{surround}$. The red shading indicates parameter ranges for which the surrounding is bistable (seizure and background states coexist). **(C)** Two dimensional parameter scans of seizure onset amplitude for the parameters of percentage of minicolumns in microseizure activity, and number of patches. Top scan (green frame) is shown for a constant value of surround excitability ($P_{surround} = -1.5$), where the system is in a bistable state. Bottom scan (blue frame) is shown for a constant value of surround excitability ($P_{surround} = -4$), where the system is in a monostable background state. Both scans are essentially slices of the parameter space shown in (A), indicated by the green and blue frames. The top two small inset panels illustrate the concept of different number of patches for the same percentage of minicolumns in microseizure activity on the cortical sheet. Black indicates the location of the patches on the cortical sheet. The amplitude of onset is measured in arbitrary units (see Methods). The parameters used can be found in Table 1.

https://doi.org/10.1371/journal.pcbi.1005475.g002

surrounding tissue ($P_{surround}$). To study this further, we plotted the distribution of seizure onset amplitudes for each value of excitability of the surrounding tissue ($P_{surround}$) in Fig 2B. High amplitude onset patterns (with amplitudes higher than 0.5) are generally found in areas of high surround excitability ($P_{surround} > -2$). When the surround excitability drops, the onset amplitude is generally below 0.5. Indeed, the parameter region supporting high amplitude onset patterns overlaps very well with the parameter region supporting bistable dynamics (between −2.5 < $P_{surround}$ < −1.1), i.e. the seizure and background states coexist in this parameter region (Fig 2B). The total percentages of minicolumns with microseizure activity, and the number of patches they form, introduce some variation in the onset amplitude (Fig 2B). Thus we decided to study their effect in two scenarios of the surround excitability: one where the surround is monostable in the background state, and one where the surround is bistable.

For $P_{surround}$ = −1.5 (Fig 2C, top, green frame), the surrounding is in the bistable state. An increase in the total percentage of minicolumns with microseizure activity would lead to a faster recruitment of the surrounding. Therefore, we expect the increase in the total percentage of minicolumns with microseizure activity to generally lead to an increase in onset amplitude, which is exactly the case (onset amplitude increases from about 1 to 2.4). By the same argument, if the minicolumns with microseizure activity are arranged into more patches, this essentially means multiple initiation points for recruitment. Hence we expect the total recruitment to be faster, and the onset amplitude to be higher, which again is confirmed in the scan (onset amplitude increases by about 0.6 on average with increasing number of patches). For $P_{surround}$ = −4 (Fig 2C, bottom, blue frame), the surrounding is in the monostable background state. In this scenario, recruitment of the surrounding is not induced easily. At the initial phase of the seizure, merely the microdomain oscillatory activity is detected on the ECoG. Therefore, the increase in total percentage of minicolumns with microseizure activity leads to a slight increase in onset amplitude (by about 1.5 when only one patch is present). Conversely, if the minicolumns with microseizure activity are arranged into several patches, their oscillatory activity is not coordinated and on average would appear with smaller amplitude than a single patch. Note that both effects are smaller in amplitude than the described effects for the bistable scenario, i.e. Fig 2C top and bottom figures have different colour axis ranges. Finally, as already shown in Fig 1, the low amplitude onset is generally associated with a slightly higher frequency than the high amplitude onset patterns. This can be seen again in Fig 2B. In the parameter region of increased excitability, the frequency of onset is significantly decreased. Again, this is due to the intrinsic properties of oscillatory domains. Large contiguous domains of seizure activity show a lower oscillation frequency in the model than smaller domains.

We conclude that overall the high amplitude slow (HAS) onset pattern is mostly strongly associated with an increased excitability of the surround leading to bistability of background and seizure state, and the low amplitude fast (LAF) onset pattern occurs in the monostable background setting. The triggers of the seizure onset in both cases are localised microdomains displaying increases in activity. The amount of microdomains and the spatial arrangement of these can influence the onset amplitude to some degree. However, the dominant parameter for the seizure onset amplitude is the excitability level of the surrounding tissue.

### Effect of surgical resection for different onset patterns

To investigate the consequence of the insight that altered surround excitability might underlie different onset patterns; we simulate the impact of surgical resections in the model for both onset patterns. This investigation is motivated by the clinical indication that the low amplitude onset pattern might be associated with a better post-operative outcome in focal seizures [9, 10, 12, 14, 16, 18]. In Fig 3 we show the evolution of both onset patterns before and after a simulated surgery. The *in silico* surgery in this case is performed by resecting 20% of the cortical sheet, which includes the area from where the onset starts. In this case it is an area of particularly high microseizure density. In order to ensure a fair comparison, we have chosen the same location and dynamics for the microseizure domains in both onset patterns. The only difference between the two scenarios is the surround excitability. In the case of the high amplitude

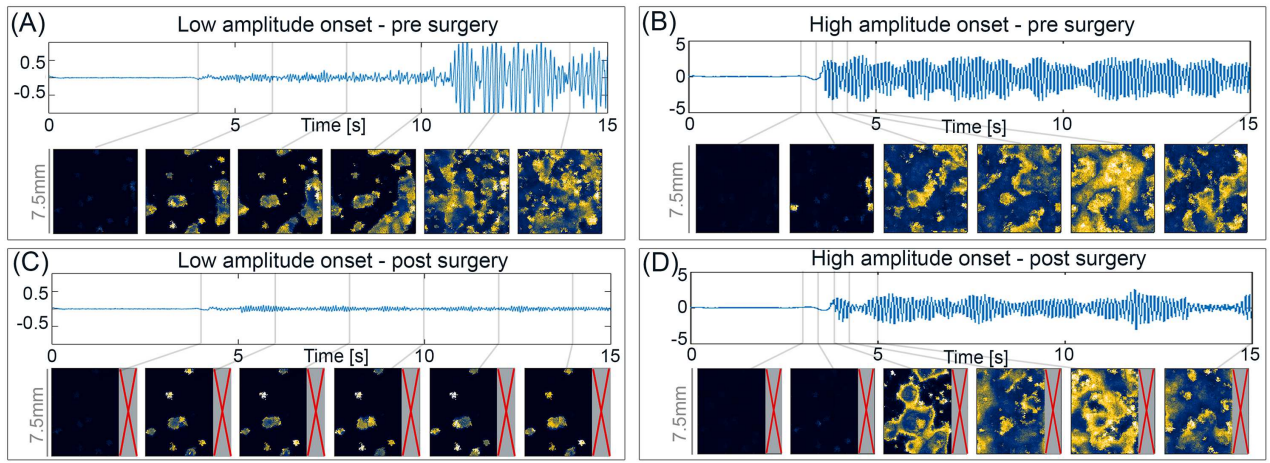

**Fig 3. Simulated surgery for different seizure onset patterns. (A)** Simulation of a low amplitude onset, similar to previously shown. The surround is monostable in the background state $P_{surround} = -4$. Snapshots of the sheet were taken at the time points indicated by the grey bars in the time series. **(B)** Simulation of a high amplitude onset, in this case using exactly the same microdomain location and dynamics as in (A). The surround is bistable with $P_{surround} = -1.5$. **(C)** Same simulation as (A), only without the resected tissue (marked by the grey box with the red cross). **(D)** Same simulation as in (B), only without the resected tissue (marked by the grey box with the red cross). The parameters used can be found in Table 1.

https://doi.org/10.1371/journal.pcbi.1005475.g003

onset, the surrounding is chosen to be bistable. In the scenario of low amplitude onset, the surrounding is chosen to be monostable in the background state.

Fig 3A and 3B show both onset patterns as we would expect before the simulated resection. After the simulated surgery, in the case of the LAF onset pattern (Fig 3C), seizure activity is still seen in some microdomains, but these no longer recruit the surrounding. Only a very low amplitude fast oscillation remains visible on the ECoG. In case of complete resection of all microdomains, the tissue would not show any remaining seizure activity. However, for the HAS patterns (Fig 3D), the simulated surgery removed the original trigger of the seizure, but does not stop full recruitment of the remaining tissue, as other microdomains are now equally able to trigger recruitment. We conclude that in our model, the high amplitude onset pattern is also associated with a worse surgical outcome. This is due to the increased excitability of the surround leading to bistability, which after surgery enables recruitment from alternative triggers. In the low amplitude onset pattern, the increased excitability in the surround is lacking. Therefore, the removal of the seizure onset zone is sufficient to prevent full recruitment. Our result is to be seen as supportive evidence that LFA might be associated with better surgical outcome. However, as described in the introduction, the clinical picture is still unclear due to the differing definitions of onset patterns, and subtypes of epilepsy investigated.

### Alternative treatment options for high amplitude onset patterns

Finally, we suggest a perspective for possible alternative treatment options, especially in the case of seizures involving a high excitability level in the surround. As we have demonstrated, surgical removal might not be sufficient as a treatment for patients suffering from this type of seizure onset. The main challenge is the "global" bistability, where the seizure state coexists with the background state. This might not be easily treated by removal of tissue, but rather requires a "global" treatment at the correct time. (See Discussion for more details on the possible spatial extent of such bistabilities). Future studies have to elucidate the exact extent of such areas of increased excitability. If the spatio-temporal extent of such areas is known, this

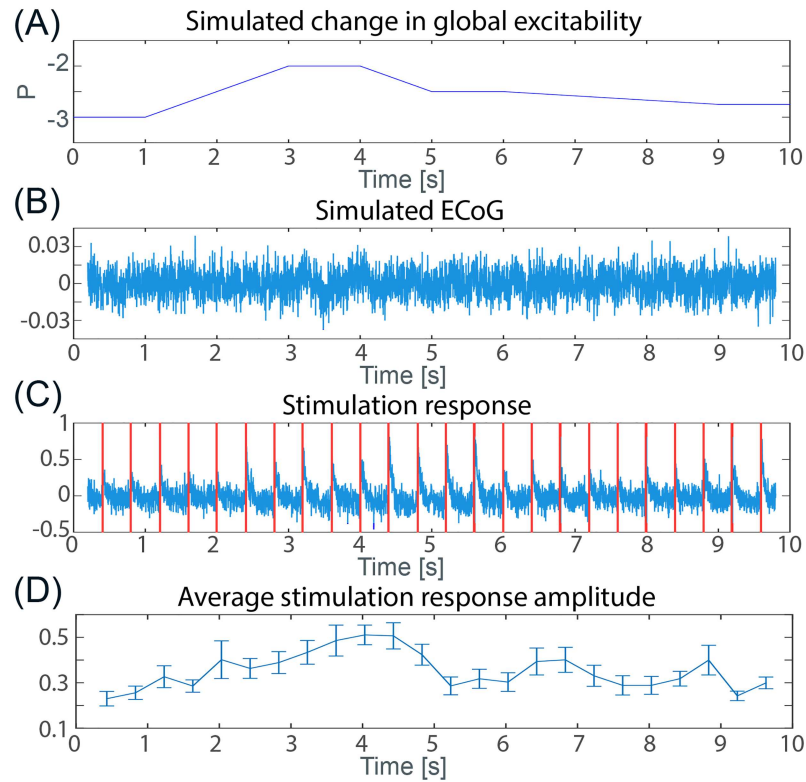

**Fig 4. Simulated microstimulation to track global changes in excitability.** **(A)** Varying model input *P* over time in all spatial locations, to simulate changing levels of global excitability. **(B)** Resulting mean-field recording (ECoG). **(C)** Stimulation responses at one of the 9 stimulation sites. Red vertical lines mark stimulation time. Stimulation duration was 6 ms. (See Methods for more details). **(D)** Maximal stimulation response averaged across all 9 stimulation locations plotted over time. Error bars indicate the standard error, giving an estimated range for the actual sample mean (if more stimulation electrodes were used). Pearson correlation coefficient of (A) vs. (D): 0.8246, $p < 0.001$. The parameters used can be found in Table 1.

https://doi.org/10.1371/journal.pcbi.1005475.g004

knowledge can be used to either surgically treat the affected tissue (e.g. multiple subpial transections), or for example to inform therapeutic brain stimulation devices when and where to stimulate.

One inherent problem of inferring where and when these increased cortical excitabilities occur is that these changes might not be observable by passive recordings (of ECoG, LFP, or even neuronal firing). This is because the increased cortical excitability can create bistable states, but the pathological abnormality is latent and invisible electrophysiologically. Hence, in order to infer when and where these bistabilities might exist, it is necessary to track changes in excitability spatio-temporally. One way to achieve this is by active probing. Fig 4 shows a proof of principle simulation, where we track the changing global excitability level over time using stimulation responses.

We simulated an entire cortical sheet undergoing changes in excitability levels over time (Fig 4A). The resulting ECoG recording is shown in Fig 4B. No obvious observable difference can be found for different levels of excitability. We therefore analysed the stimulation response amplitudes, as it is an established clinical way of measuring excitability levels [15, 31]. In 9 locations on the sheet we simulated stimulation (affecting one local minicolumn) using a transient increased input to the excitatory populations (parameter P) for 6 ms. The result of one of the 9 stimulation sites is shown in Fig 4C. We then measured the maximal stimulation response, and averaged this over the 9 stimulation sites, giving us an average stimulation amplitude (Fig 4D). This average stimulation amplitude was found to track the underlying changes in the excitability levels (Pearson correlation coefficient: 0.8246, $p < 0.001$).

## Discussion

In summary, we conclude that the two focal seizure onset patterns are distinguished by differing tissue excitability in our model. The high amplitude onset pattern is associated with an increased surround excitability that creates the coexistence (bistability) of a seizure state in the surrounding tissue. Conversely, the low amplitude onset pattern requires domains displaying localised seizure activity, which are embedded in a surrounding that is not bistable. As a direct consequence of this, the model shows that local removal of (epileptogenic) tissue can prevent recruitment and a full-blown seizure in low amplitude onset patterns, but not necessarily in high amplitude patterns. As an outlook, and to address the question of what alternative to removal of tissue one can pursue to prevent seizures in the case of a bistable surround, we propose the idea of tracking surround excitability levels to enable the operation of targeted intervention devices.

### Modelling of onset patterns

In our model we were able to characterize two main types of onset pattern, the low amplitude fast pattern, and the high amplitude slow pattern. These two patterns agree qualitatively with seizure onset patterns described in the clinical literature in their initiation and evolution of the activity amplitude, and also to some extent in frequency of the seizure activity. However, the clinical literature highlights a range of different onset pattern morphologies with a high amplitude at onset, e.g. high amplitude spikes, sharp waves, or high amplitude spike-and-wave activity, etc. [17, 18]. In our model, we did not investigate the possibility of generating the different subtypes of a high-amplitude onset pattern, as we were focusing on the difference in onset pattern introduced by a bistable surround. Nevertheless, it is worth mentioning that this is not an intrinsic limitation of the model, as for example we and others have shown in earlier studies that more complex seizure waveforms can be generated even in "simple" neural population models such as the one used here (e.g. [28, 29, 32, 33]). Hence it is not inconceivable that more complex onset patterns could be found in the model with the right parameter (time scale) settings. We leave this investigation for a future study.

In terms of seizure onset, the diffuse electrodecrement or attenuation onset pattern deserves attention, as it has also been shown to correlate with a worse surgical outcome [16, 18]. Currently, to our knowledge, neither our model, nor any other model has provided a mechanistic insight regarding this pattern. Particularly, as it may not only be a seizure onset pattern, but can occur in the middle of a seizure [34]. The spatial scale of this pattern is also generally diffuse, often involving most of the recording electrodes [18], making it a difficult event to study for our model. However, we speculate that the diffuse nature may indicate a wide-ranging abnormal surround. If that is the case, the abnormal surround may be the real driver behind the seizures, and local removal of the apparent seizure onset zone may not be indicated. To extend our understanding of this pattern further, a model of different spatial scale (whole brain scale, e.g. [35–37], or at the scale of an ECoG grid, e.g. [38–40]) is perhaps required, potentially also requiring subcortical structures (e.g. [41, 42]).

We noted that the faster frequency for the low amplitude onset pattern is an intrinsic property of independently oscillating patches of tissue in the model. This is not only an intriguing phenomenon in the model, which deserves further investigations, but might also suggest an alternative mechanism by which frequency changes during a seizure can be explained. The exact frequency range for low amplitude fast activity at seizure onset is so far inconsistently defined between studies. For example some studies call all frequencies above 8 Hz as fast activity [16], while other studies use a cutoff frequency of 13 Hz [10], or even higher [14]. However, it appears that all studies refer to oscillations in the alpha/beta band. In that regard it might be interesting to clarify clinically 1) the amplitude and frequency range of these fast onset activities, 2) their relationship to the background activity, and 3) how they evolve over time, to enable future modelling efforts in this direction.

It is also an open question how these fast activities in the beta band relate to high-frequency oscillations (HFOs) at seizure onset (gamma range or beyond, see e.g. [32]). Multiple suggestions for the cellular mechanisms of HFOs exist, and their role in epilepsy is also being discussed [43]. High frequency oscillations phase-locked to the lower frequency band (1-25 Hz) at seizure onset have been proposed recently to be related to the multi-unit activity of neural populations [44, 45]. In an adequate translation of our model (converting the relative firing rates to multi unit activity and their HFO signatures), these phase-locked HFOs could also be observed in our model output. This is because the firing rates increase at seizure start, which is directly translated to the LFPs we show in our figures. In our model an increase in HFO would thus be observed in both onset patterns, perhaps with a lower power in the low amplitude onset pattern. Perucca *et al.* indeed observe that the HFO rate increases in general at seizure onset, independent of the onset pattern [17]. However, this will deserve a closer examination and direct simulation in order to confirm that our model may be able to show HFOs at seizure onset. Furthermore, when distinguishing between ripples (80-200 Hz) and fast ripples (200-500 Hz) in the HFOs, Levesque *et al.* show that the distribution of the two types of HFOs differs between different types of seizure onset in a rat model of temporal lobe epilepsy [46]. The authors find a higher ripple rate associated with the LAF pattern, and a higher fast ripple rate associated with the HAS pattern. Future studies (possibly with more detailed neuronal models) will show if these observations can be related to our hypothesis of differences in surround excitability and onset mechanism.

The choice of our spatio-temporal model for this study is crucial, as we needed to investigate the spatial interaction of epileptogenic patches with their surround. This was possible, as our previous study [21] showed a natural way of understanding surround excitability/bistability and the incorporation of epileptogenic patches. Furthermore, the model is computationally efficient given the volume of tissue simulated, and the level of abstraction of the model has proven sufficient in capturing the onset patterns in their most obvious qualitative features. However, we acknowledge that other details may prove important to capture more details and mechanisms of onset patterns in future studies. There is still much work to do in future to achieve a full understanding of what generates different onset patterns spatio-temporally, and our work can serve as a starting point for future *in silico* approaches, as it naturally links key concepts (tissue excitability, recruitment, and core/surround) with observations of onset patterns.

### Spatio-temporal extent of increase in excitability

So far, we have only commented briefly on the possible spatial extent of the surround with increased cortical excitability (which we termed "global" increase in excitability). Badawy *et al.* measured increases in cortical excitability by the motor cortex threshold, and showed that such increases can be detected even in focal seizures with the seizure onset zone (SOZ) not in the motor cortex [2]. Hence, this suggests that the spatial extent of areas with increased excitability may involve widespread networks, perhaps even an entire hemisphere. However, [15] showed that the CCEP response amplitude was significantly lower in electrodes far away from the SOZ, suggesting that the area of increased excitability could be limited to the coverage of an ECoG grid. Furthermore, [47] showed that delayed responses to single pulse electrical stimulation are mainly seen in the SOZ, suggesting that this area has altered excitability levels. Further studies are required to explicitly map the spatio-temporal evolutions of the (abnormal) shifts in excitability on the cortex.

Interestingly, Enatsu *et al.* also found that in both onset patterns, the excitability of the SOZ is increased compared to a location of the ECoG that is far away from the SOZ [15]. This suggests that an increase in the excitability of the surround might occur in both onset patterns. This is also in agreement with findings that motor thresholds are consistently altered in focal seizure patients and not just a subset of patients [2]. In the model this observation is also not surprising. Tissue with localised patches of microseizure activity would also display a slight increase in stimulation response, but not as much as compared to the bistable surround. However, this also needs to be accompanied by improvements in clinical recordings to provide a higher resolution spatio-temporal clinical picture, over more extended areas. Additionally, computational models that span the spatial scales from minicolumns to an entire ECoG grid may be beneficial to tease apart the different spatial scales of excitation that could occur.

The source and mechanism causing the increased surround excitability is also unclear. In both patients with acquired structural epilepsies and their siblings, an increase in motor cortex excitability could be demonstrated [48]. This suggests that the increased cortical excitability is not necessarily a pathological feature, but rather to be understood as an enabling, pro-ictal condition. The study also suggests that genetic factors might underlie the increased excitability.

In cases of lesional epilepsy, tumors, or congenital dysplastic lesions, the surrounding tissue is generally also considered abnormal, and is therefore also resected. The increased surgical success rates in such cases [49] indicate that resection of the surround is a feasible and working strategy. However, it is also conceivable that the tissue around a tumor are in fact the seizure core, and the surround as we understood it in our model is the wide brain network the tumor and surrounding tissue are embedded in. Again, this highlights that the spatial scale at which to consider 'surround' and 'core', and the meaning of 'increased excitability' need more precise clinical and theoretical definitions. In this context, we wish to highlight that our results do not necessarily indicate a bigger resection area (i.e. also remove the increased surround excitability). Rather, our results suggest that some patients may have an enabling surround (of unknown spatial extent), predisposing them to generate seizures more easily. The increased surround excitability is to be understood as a pro-ictal condition. Such a condition on its own would not generate seizures, but needs provocation or driving from an epileptic generator. In patients who experience a rapid recurrence of their seizures soon after surgery, such epileptic provocations or generators may already exist (e.g. pre-existing micro-seizure domains). In other cases, there may be an early post-surgical improvement, but relapse over a longer period of time. This latter case may occur if a new epileptic generator forms slowly over time, perhaps by mechanisms of plasticity [50].

### Model prediction and future validation

To further validate our model hypotheses, we derived two predictions to compare to clinical literature. Firstly, the model predicts an increased excitability in and around the seizure onset zone (SOZ) for high amplitude onset pattern. Indeed Enatsu *et al.* showed that the stimulation response amplitude (after CCEP) is higher in the seizure onset zone (SOZ) of high amplitude onset seizures compared to low amplitude onset seizures [15]. Our model further predicts that in the low amplitude fast onset pattern, recruitment is only possible given a distributed network of patches of microseizure activity. This means that for seizure initiation, (micro-)seizure activity at multiple sites must be visible. We restricted our simulation to the spatial extent of a single ECoG electrode, but it is conceivable that such networks of patches can extend to multiple electrodes. Hence we would expect the observation of a wider seizure onset zone (i.e. more electrodes to show seizure activity at onset). This is indeed also the case, as reported by [17].

Further support for our model is provided by a recent study, which used microwires in conjunction with depth electrodes in mesial temporal lobe structures to record temporal lobe seizures [51]. Their findings confirm the spatio-temporal patterns observed in our model: low amplitude fast onset patterns are shown to begin with small patches of seizure activity, which slowly grow in size and coalesce in the initial phase of the seizure. However, in order to fully test the validity of the model finding, high-resolution spatio-temporal recordings with a significant spatial coverage (e.g. high density electrode arrays [52]) would be required to compare the evolution of the seizure in patients to our model simulations. Additionally, to test for increased excitability levels, spatially, and temporally, an active stimulation paradigm might be required. Future studies should systematically test the excitability levels at all spatial locations covered by ECoG, possibly at different time points, (i) to elucidate if the cortical excitability around the SOZ is increased in high amplitude compared to low amplitude onset seizure, and (ii) to determine the spatial extent of such an increased excitability.

## Outlook on alternative treatment strategies

Finally, we proposed the perspective of a spatio-temporal monitoring system for cortical excitability. Such a system would offer the possibility of alternative treatment strategies to surgical removal of tissue. In particular, a closed-loop control system can use the information on cortical excitability to deliver spatio-temporally precise control stimulation. Various control frameworks exist already, including electric [53–57], magnetic [58], and optogenetic [59] control. We also summarised several theoretical investigations in our recent review [60], and also see [61]. The key contribution of our work towards development of such closed-loop control systems is that we highlight the need for a spatio-temporal controller. Much of the recent attention has been focused on temporal control, but the spatial aspect has been largely neglected. We demonstrated here that the distinction of spatial territories into seizure core and surround is crucial to understand seizure onset. Furthermore, we demonstrated that only controlling the core might not be sufficient in some types of seizures, but controlling the surround could be as important.

## Conclusion

Previous studies have demonstrated that the spatial distinction of a seizure core and a surround is a useful concept in the theoretical [21] and the clinical [25, 44, 45, 62] context. Using this concept allowed us to demonstrate that the excitability level of the surround is a crucial factor in determining the onset pattern, and the pathological abnormality in the model. We also suggested further studies to be performed to test our model hypotheses; and the validity of the concepts proposed here. If validated, our model essentially suggests that patients with different onset patterns should be treated differently. In the case of low amplitude fast onset patterns, a traditional resection of the SOZ should lead to seizure freedom. In the case of high amplitude onset patterns, we suggest the control of the excitability in the surround to be crucial.

**Table 1. Parameters used for the figures.** All parameters remain the same as in our previous work [21], except for the parameters specifically listed below.

| Figure | $P_{surround}$ | $P_{patch}$ | Number of patches | Percentage of hyperactive minicolumns |
|---|---|---|---|---|
| 1 (a) | -2.5 | ramped from -2.5 to 2 within 3 s | 15 | 5% |
| 1 (b) | -1.5 | ramped from -1.5 to 1 within 3 s | 1 | 5% |
| 2 | scanned | ramped to 1 within 3 s | scanned | scanned |
| 3 (a,c) | -2.5 | ramped to 1 within 3 s | 15 | 5% |
| 3 (b,d) | -1.5 | ramped to 1 within 3 s | 15 | 5% |
| 4 | varies over time | none used | 0 | 0 |

https://doi.org/10.1371/journal.pcbi.1005475.t001

## Methods

### Model of a cortical sheet

The model used here has been introduced in detail in our previous publication [21], and we use the same basic settings and parameter values as in the previous publication (specific changes in parameters are shown in Table 1). The MATLAB code for the model is also published on ModelDB (ID: 155565) alongside the previous publication.

The basic unit of our model is a cortical minicolumn, which is modelled by a Wilson-Cowan unit (Wilson and Cowan, 1973) consisting of an excitatory and an inhibitory neural population ($E$ and $I$). It assumes that the $E$ and $I$ populations interact with each other (Fig 5A), and the this interaction influences the firing activity of the target population. This established model has been shown to be able to capture some coarse-grained dynamics of neural populations [63–66], and is at the same time quick to simulate. Importantly, we demonstrated in our previous work [21] that this model offers a good trade-off between level of abstraction while still capturing key seizure dynamics to study seizure onset mechanisms. It allows for the simulation of spatio-temporal patterns of seizure activity, and the analysis of spatial tissue heterogeneities and mesoscopic connectivity in that context.

The equations to simulate a single minicolumn is, exactly as in our previous study [21], a simplified Wilson-Cowan unit given in Eq 1:

$$\tau_E \cdot \frac{dE}{dt} = -E + Sigm(C_{E \to E} \cdot E + C_{I \to E} \cdot I + P + A_s \cdot S(t))$$
$$\tau_I \cdot \frac{dI}{dt} = -I + Sigm(C_{E \to I} \cdot E + C_{I \to I} \cdot I + Q),$$

(1)

where $E$ is the fractional firing activity in the excitatory population; $I$ is the fractional firing activity in the inhibitory population. $P$ and $Q$ model the baseline activation of the $E$ and $I$ populations (or sometimes also referred to as the basal input level). $S(t)$ is a noise input to $E$ to reflect noise, or input from other brain areas that is uncorrelated to the dynamics of local interest. $A_s$ is the coupling strength of the noise input. The connectivity constants $C_{i \to j}$ (with $i,j$ = $E$ or $I$) determine the coupling strength between the populations, which essentially indicate how input from the source population is interpreted by its target population (see Fig 5). E.g. $C_{E \to I}$ controls how input from the $E$ population influences the $I$ population.

$Sigm(\ldots)$ is a sigmoid function, which is derived from a distribution of firing thresholds in the underlying neural population [67]. It is defined as $Sigm(x) = \frac{1}{1+exp(-a(x-\theta))}$, where $a$ is the steepness of the sigmoid and $\theta$ is the offset (in $x$) of the sigmoid. We fix the sigmoid parameters ($a = 1, \theta = 4$) following previous work [33], as variations in the other parameters $P,Q,C$ effectively result in a change of the sigmoid shape.

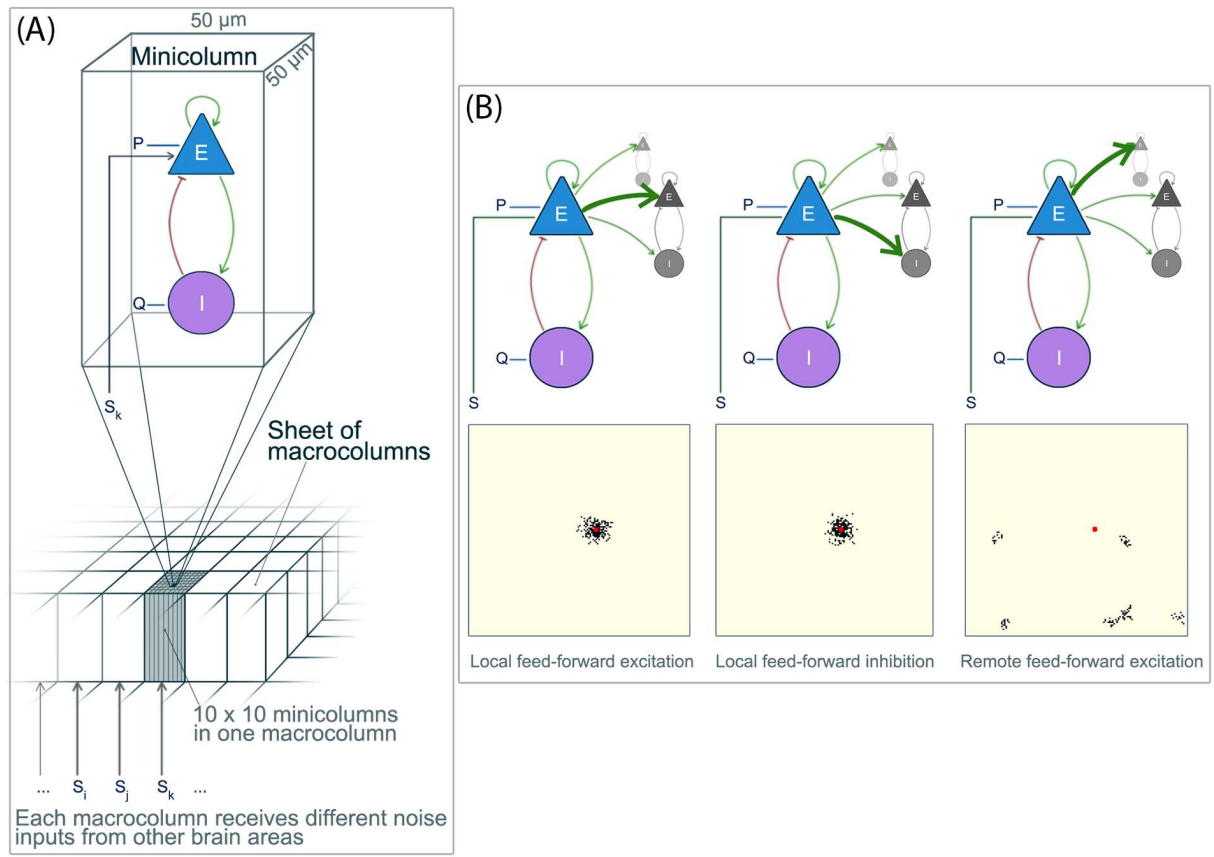

**Fig 5. Schematic illustration of the simulated cortical sheet. (A)** The cortical sheet consists of cortical minicolumns, each modelled by a Wilson-Cowan unit consisting of an excitatory (E) and inhibitory (I) population. The dynamics of a single minicolumn is determined by the interaction of the E and I populations, together with the baseline excitation (P and Q). In this study we simulated a cortical sheet consisting of 150x150 minicolumns. Every 10x10 minicolumns make up a macrocolumn, which receive the same subcortical noise input (S). **(B)** On a cortical sheet the minicolumns additionally interact laterally with other minicolumns. The top row illustrates schematically the three different types of feed-forward interactions that are included in our model. The bottom row shows one exemplary minicolumn (red) and its outgoing connection targets (black) on the simulated cortical sheet. Figure modified from [21].

https://doi.org/10.1371/journal.pcbi.1005475.g005

To form a cortical sheet, we concatenate 150 by 150 minicolumns (Fig 5A), and couple them to each other. Note that the full system is also described by Eq 1, if we understand *E*,*I*,*P*,*Q* and *τ* as vectors, and the connectivity *C* as matrices. The spatial scale of the cortical sheet is derived from the assumption that each minicolumn is about $50\mu m \times 50\mu m$ in size [68]. A macrocolumn is then formed by $10 \times 10$ minicolumns, which agrees with the size suggested in [69]. The lateral connections between the minicolumns include local feed-forward excitation and inhibition, as well as remote feed-forward excitation (see Fig 5B for a schematic). The projection targets of local feed-forward excitation and inhibition are chosen with a certain probability that falls off with distance, as is shown in slice studies (e.g. [70]), and as is adopted by most modelling studies (e.g. [71, 72]). We chose parameters for cortical connectivity following the suggestions in [73], which are derived from tract tracing experiments in human cortical tissue. Specifically, [73] also suggest to include the so-called remote feed-forward excitation. This is a mesoscopic type of connection established by principle neurons, which target clusters of cells which can be several millimetres away (see Fig 5B, right column for an example). The exact algorithm for finding the connectivity is described in the supplementary information of our previous work [21], following [73], and we also provided these connectivity matrices with our code.

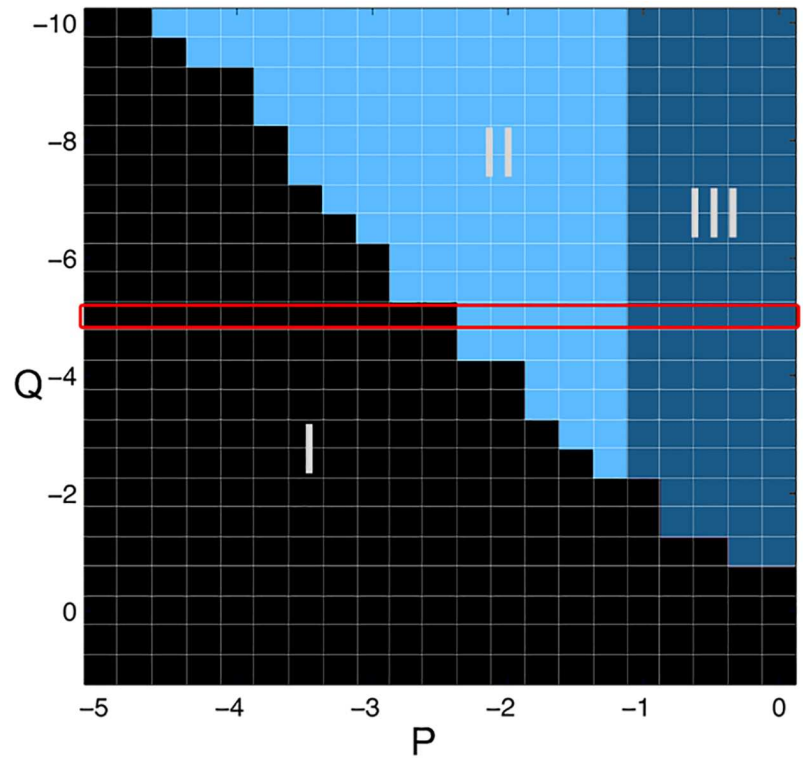

**Fig 6. Parameter space of the model in terms of model dynamics.** I: Region of background activity, where we find a low rate of unstructured firing. II: Bistability between background and seizure state. III: Region of seizure activity, where we find a high rate of firing that is oscillating. The red frame highlights the parameter area we operate in for the subsequent figures. Figure modified from [21].

https://doi.org/10.1371/journal.pcbi.1005475.g006

Finally, the noise input is structured at the cortical sheet level, such that a macrocolumn (10 by 10 minicolumns) receives the same $S(t)$ [68, 74]. We used noise values drawn from a standard normal distribution. The effective noise coupling strength is set to $A_s = 1$. In this setting the system is not entirely dominated by the noise input but the noise influences the deterministic dynamics. Simulations of the system used a Euler-Maruyama fixed step solver, with a stepsize of 2 ms. Qualitatively equivalent results are found for smaller stepsizes. Fig 5 schematically summarises the model. Supplementary S1 Text additionally shows some simulation results for the deterministic system (i.e. without noise input) for comparison and completeness.

### Dynamics of the simulated cortical sheet

In our previous analysis [21] we demonstrated that the dynamics of the model cortical sheet can be characterised by distinct states. The sheet can exist in a background state, with little activity, where low amplitude noise dominates (region I in Fig 6). The second state is an oscillatory activity state (region III in Fig 6). We identify this oscillatory state as the seizure state (see [21]). Interestingly, a parameter region where both the background and seizure state coexist can also be identified (region II in Fig 6). This coexistence of states is termed "bistability". (We call the case of only one state existing "monostability"). Here we only show the parameter space for the two input parameters ($P$ and $Q$, which essentially model the level of baseline activation of the $E$ and $I$ populations, respectively). Equivalent states can be found for other parameters. For the majority of the manuscript, we operate in the parameter region outlined in red in Fig 6.

Importantly, we demonstrated previously that transitions to seizures can be triggered by a short transient input/stimulus when the simulated cortical sheet is in the bistable regime [21]. Such a bistable mechanism of seizure onset has also been proposed by other theoretical studies [26, 75]. However, we also show in our previous study that even in the monostable background state, it is possible to induce a transition to the seizure state in the whole sheet, when microdomains with autonomous seizure activity are embedded in the cortical sheet [21] (inspired by the clinical observations of microseizures [3]). In other words, seizures cannot only be provoked by transient stimuli in the bistable regime, but can also slowly penetrate into "healthy" monostable background tissue. We also presented the behaviour of the system in a dynamical systems context in S1 Text, to aid the understanding of our work in the theoretical domain.

We also use the term "excitability" throughout the manuscript. We will use this term to describe the proximity to the monostable seizure state in parameter space. In other words, how close the current parameter setting is to the dark blue region (III) in parameter space in Fig 6. This definition is intuitive in the context of our paper, because we showed in our previous work [21] that the proximity to the monostable seizure state directly relates to how "easy" it is to provoke seizure activity from perturbation stimulation. In our case, "easy" refers to how big the stimulus has to be spatially, and how strongly it activates the populations.

Increasing excitability throughout the manuscript usually refers to increasing $P$ if not mentioned otherwise. However, $P$ is not the only parameter that can lead to a change of excitability. For instance a decrease in $Q$ would have the same effect as it moves the system closer to the seizure state. Indeed our results still hold when using a decrease in $Q$ instead of an increase in $P$ (see for example S1 Fig). In addition, we tested our results in this manuscript with regards to robustness to changes in conduction delays and boundary effect, we found no qualitative difference, and all conclusions still hold.

### Simulating microdomains with microseizure activity

In order to simulate seizure onset, we focus on mechanisms that are triggered, or caused by small patches of localised seizure activity (termed class II and class III onset in our previous work [21]). These small patches will also be referred to as microdomains, and seizure activity restricted to such small patches of tissue will be termed microseizures [3]. A patch refers to a group of spatially contiguous minicolumns. To model microseizures, we put a patch in a monostable seizure state (in our case, we gradually changed $P_{patch}$ over 3 seconds to $P_{patch} = 1$). $P$ for the surrounding cortical sheet is termed $P_{surround}$, which is usually either set to −2.5 to make the surrounding monostable, or −1.5 to make the surrounding bistable.

To model different onset patterns, we scanned the number and spatial arrangement of the patches of microseizure activity. The spatial arrangement of patches is chosen randomly. Hence we scan 10 different configurations with the same algorithm, and report average results of seizure onset amplitude in our scans.

In the case of multiple patches in space, we assume that they do not all show seizure activity at once, but rather transition to the seizure state with a 50 millisecond gap between each patch. This is to ensure that we are not merely detecting our effects due to synchronisation effects following simultaneous activation. We also scanned the 'onset time gap' (between 6 and 100 millisecond, and also random time gaps) and found no qualitative difference in our results.

To ensure reproducibility, we shall also make the code to generate low and high amplitude onset patterns available on ModelDB (ID: 226074). The code includes the details described here.

### Parameter settings

Table 1 shows all the parameter we used for the figures in the Result section.

### Simulating LFP and ECoG, detecting seizure onset

As in our previous work, we model the local field potential of a minicolumn by a weighted sum of its excitatory and inhibitory activity, and its noise input. This measure is an approximation of the post-synaptic potential in the pyramidal (excitatory) population. To model the recording of an ECoG that would lie over the simulated sheet, we used the average of all minicolumn LFPs, as an approximation. This average signal is then high pass filtered at 1 Hz, as clinical recordings do not usually include the DC component. (Our model actually shows a DC shift at seizure onset, but the details are not shown here. More related information can be found in a recent publication [29]).

We used a series of different algorithms (amplitude deviation detection based on the stable background amplitude; DC shift detection in the unfiltered time series; and deviation of steepness of time series from the background) to detect seizure onset from the simulated ECoG signal. All algorithms showed qualitatively similar results.

### Simulating surgery

We simulated surgical removal of cortical tissue by removing all connections to and from the resected area. The resected area in this case is a strip of the sheet, 150 × 30 minicolumns in size.

### Simulating microstimulation

To simulate stimulations in one minicolumn, we raised their input level *P* by 5 parameter units for 6 milliseconds. I.e. the target minicolumn receives a transient input via a short pulse to the parameter *P*. In order to gauge the spatial variance, we simulate the stimulation to different positions on the cortical sheet, we target 9 minicolumns in total. Their locations are arranged on a 3 by 3 grid on the cortical sheet, in positions 25, 75, and 125 in horizontal and vertical dimensions, respectively. The stimuli are delivered successively at spatially distant sites to minimise interference. This means that only one location is stimulated at any given time. The stimuli between sites are 6 milliseconds apart from each other. This is to avoid joint activation of all sites.

The stimulation responses are measured at each stimulation location. For this, we simply read out the LFP of the stimulated minicolumn. A moving window average is applied (window length of 20 milliseconds) to smooth signal, and the maximum value is taken as the response amplitude from the smoothed signal at each stimulation location. The average response amplitude at any time is simply the mean of the response amplitudes of all 9 locations.

This stimulation procedure is then repeated every 400 milliseconds, meaning that every 400 milliseconds, we derive a measure of average response amplitude over the whole simulated cortical sheet. This protocol has not been developed to be optimal, but is intended to demonstrate the principle of how excitability levels could be tracked using small stimuli.

## Supporting information

**S1 Fig. Seizure onset amplitude depending on $Q_{surround}$, number of patches, and percentage of hyperactive minicolumns.** This is the equivalent figure to Fig 2 (a), only using *Q* as the parameter to change excitability. Seizure onset amplitude is, again, shown in the size of the marker.
(PNG)

**S1 Text. Deterministic behaviour of the model to demonstrate some theoretical aspects underlying the dynamics.** We show the bifurcation behaviour in the deterministic system, and specifically focus on the monostable state, and under what conditions microdomains can still recruit their surrounding tissue.
(PDF)

## Acknowledgments

We thank the CANDO team (particularly Andrew Jackson), Piotr Suffczynski, Louis Lemieux, and Gerold Baier for fruitful discussions.

## Author Contributions

**Conceptualization:** YW.

**Data curation:** YW.

**Formal analysis:** YW.

**Funding acquisition:** AJT MK.

**Investigation:** YW PNT.

**Methodology:** YW.

**Resources:** YW PNT MK.

**Software:** YW PNT.

**Validation:** YW.

**Visualization:** YW PNT.

**Writing – original draft:** YW PNT.

**Writing – review & editing:** YW AJT PNT AV GA MK.

PLOS COMPUTATIONAL BIOLOGY

Mechanisms underlying different onset patterns of focal seizures